%% file: main_plain.tex
\documentclass[9pt,twocolumn]{article}

\usepackage[top=0.75in, bottom=1in, left=0.625in, right=0.625in]{geometry}

\usepackage{amsmath, amssymb}
\usepackage{algorithm}
\usepackage{algpseudocode}
\usepackage{graphicx}
\usepackage{hyperref}
\usepackage[center]{caption}
\usepackage{todonotes}

\newcommand{\precision}{\text{Precision}}
\newcommand{\recall}{\text{Recall}}
\newcommand{\centercell}[1]{\multicolumn{1}{c}{#1}}
\newcommand{\p}{\mathbf{P}}

\title{Supervised Classification of LEO Debris Families Using Multi-Set Proper Elements}
\author{Michael Ling \\ University of New South Wales, Sydney, NSW 2052, Australia \\ \texttt{michael.c.ling@student.unsw.edu.au} \\ \\ Yang Yang \\ University of New South Wales, Sydney, NSW 2052, Australia \\ \texttt{yang.yang16@unsw.edu.au}}
\date{\today}

\begin{document}

\maketitle

    \input{sections/abstract.tex}

    \input{sections/intro.tex}
    \input{sections/method.tex}

    \input{sections/results_and_discussion.tex}
    \input{sections/conclusion.tex}

\input{sections/acknowledgements.tex}

    \bibliographystyle{plain}
    \bibliography{references.bib}

\end{document}

%% file: sections/abstract.tex
\begin{abstract}
    The use of proper elements to reconnect families of satellite fragmentation
    debris has recently advanced with the emergence of machine learning techniques.
    These developments are key to improving the efficiency and effectiveness of operations in space sustainability operations, collision avoidance, and space domain awareness.
    However, a constantly evolving circumterrestrial environment may limit
    the applicability of these techniques, especially when the nonlinear thresholding
    learned by neural network models is skewed by training on outdated
    representations of debris in orbit. In this work, we devise and test a computational pipeline in a controlled environment to evaluate current
    classification techniques and suggest future directions. Synthetic
    fragmentation data from explosive breakup events were generated using a Standard Breakup Model and
    propagated under a high-fidelity dynamical model. Proper elements for each
    fragment were then extracted using an adapted algorithm for the modified
    equinoctial elements (MEE), Poincar\'e elements (PNC), and quaternion sets (QTN). 
    Previous supervised-learning approaches for debris-family classification operated only in MEE proper space. By extending proper-element extraction to PNC and QTN sets, we broaden the space of dynamical fingerprints available to the classifier.
    Fragments were classified using several neural network models trained on different combinations and modifications of the element sets to determine
    whether a pair of fragments could be attributed to the same parent.
    Crucially, we identify a fundamental limitation when applying standard quaternion sets to neural networks: the loss of orbital size information during feature normalization. We introduce an augmented representation (QTN$_p$) that explicitly restores the semi-latus rectum, which improves the accuracy from 0.31 to 0.60 compared to the standard set.
    In our synthetic Starlink-like LEO experiment, expanding and combining proper-element sets generally improves discrimination between same-parent and different-parent pairs. 
    The best model, using a joint feature set built from three distinct proper-element representations (MEE + PNC + QTN), achieves an area under the receiver operating characteristic curve of approximately 0.858 compared to 0.789 for the best MEE-only baseline, together with higher accuracy and F1. 
\end{abstract}

\vspace{1em}

%% file: sections/intro.tex
\section{INTRODUCTION}
\label{sec:intro}
Since the launch of the first artificial satellite into Earth's orbit, human
activity in the circumterrestrial environment has proliferated, driven by technological advancement and growing demand. Essential systems for
communication, defense, navigation, and observation all exploit the capabilities
of Earth-orbiting satellites to some extent. Therefore, the safety,
sustainability, and stability of this environment cannot be overstated.
However, recent activities have compromised space domain
awareness and re-established the criticality of geocentric space
\cite{Radtke2017,Long2024}.

Geocentric orbital launch traffic has risen dramatically in recent years, with the rate of object injection far surpassing the rate of end-of-life operations \cite{esa2023}.
With the advent and implementation of mega-constellations, the resident space objects (RSO) population is not expected to plateau.
Naturally, as the RSO population increases, collision risk is expected to escalate.
Furthermore, collision and explosion events, both accidental (such as battery explosions or propulsion system failures) and deliberate (such as anti-satellite tests), worsen the situation by generating persistent fragmentation clouds, contributing to the exponential behaviour known as Kessler Syndrome \cite{kessler}. 



A significant port of the debris population consists of RSOs with unknown origin
(RUOs) \cite{Wu2023}. These are objects that are tracked by space surveillance networks but have not been correlated to any known launch or breakup event in existing catalogs. The presence of RUOs complicates space traffic management and collision avoidance, as their trajectories and potential interactions with other RSOs are less predictable \cite{esa2023}. Consequently, the ability to reconnect RUOs to their parent objects through fragment family classification is vital for maintaining an accurate and comprehensive space object catalog.


The methodology of space object family classification has its roots in the celestial mechanics community, particularly in the context of asteroid family classification \cite{hirayama1927,Zappala1990,knezevic2000}. These works introduced the concept of "proper orbital elements", which are quasi-constant quantities that remain invariant under secular perturbations. Proper elements serve as dynamical fingerprints, allowing for the identification of groups of objects sharing a common origin. 

However, the application of proper elements for classifying RSOs in low Earth orbit (LEO) is a relatively recent development. While Cook \cite{Cook1966} first hinted at their potential, it was not unitl the work of  Celletti et al. \cite{Celletti2021} that the method was rigorously demonstrated for artificial satellites. 
In an idealized two-body scenario, where a satellite orbits a perfectly spherical Earth, the orbital elements, such as semi-major axis $a$, eccentricity $e$, and inclination $i$, would remain constant indefinitely. In reality, however, Earth's oblateness (captured by spherical harmonics like $J_2$), atmospheric drag, and gravitational influences from the Sun and Moon introduce perturbations that cause these classical elements to vary over time. This variability complicates the identification of fragments originating from a common parent. Proper elements address this challenge by filtering out short-period variations and isolating the secular (long-term) behavior that is more indicative of shared origin. Celletti et al. employed a Hamiltonian model and semi-analytical methods to validate their hypothesis of fragment classification using proper elements in a mixed-case scenario involving debris from both CZ-3 and Atlas V Centaur rockets \cite{Celletti2021}. Their results illustrated that osculating elements were inadequate for classification, while proper elements provided a high-confidence statistic for the same problem. Celletti and Vartolomel recently constructed a dynamics-based pipeline that combines perturbation-theory-derived proper elements, demonstrating that these quasi-invariants yield significantly more robust clustering and labelling of fragments than mean elements over multi-decade evolutions \cite{Celletti2025}.

Parallel to these dynamical approaches, the last three years have seen a surge in data-driven methodologies. Joshi et al. recently proposed a graph neural network that treats debris clouds as topologically connected graphs to distinguish between explosion and collision events \cite{Joshi2025}, while Lavezzi et al. applied Transformer architectures to raw astrometric time-series data for rapid ``early classification" \cite{Lavezzi2024}. While these geometric and temporal deep learning methods offer powerful new capabilities for short-term characterization, the reliance on stable dynamical invariants remains the standard for long-term family attribution and catalog maintenance.

Building on the dynamical approach, Wu \cite{Wu2023} developed a method to extract quasi-constant proper elements, better suited for LEO RSOs than previous asteroid-focused techniques \cite{WuThesis}. Wu combined these features with (density-based spatial clustering for applications with noise) DBSCAN clustering, but found the standard distance function \cite{Zappala1990} inadequate for separating multiple debris families. To overcome this, Wu introduced a neural network-based nonlinear distance function, enabling more general RUO classification. However, their implementation was limited to a single feature space, i.e., modified equinoctial element (MEE) space, and relied solely on difference-based inputs. This imposes an information bottleneck, potentially discarding independent dynamical signatures available in other coordinate representations. While Wu's approach advances beyond Celletti et al. \cite{Celletti2021}, it remains an open question whether richer, multi-set dynamical fingerprints are required to sufficiently separate overlapping debris families in increasingly crowded environments.

This paper addresses this gap by presenting the first systematic comparison of single, double, and triple proper-element sets for RUO classification. The major contributions of this work are summarized as follows:
\begin{itemize}
    \item We construct a modular synthetic pipeline that simulates breakup events of parent objects, propagates their fragments, extracts proper elements, and classifies the fragments into families.
    \item We propose a state-complete input representation that preserves absolute proper elements for both fragments in a pair, allowing the neural network to learn effective separation relations rather than relying on hand-crafted difference features.
    \item We extend prior MEE-only work by implementing proper-element extraction for two additional element sets, i.e., Poincar\'e (PNC) and quaternion (QTN), and systematically compare single-set, double-set, and triple-set combinations. This provides quantitative design guidance, in the specific LEO-like scenario considered here, on how element-set choices affect accuracy, F1-score, and the area under the receiver operating characteristic curve (ROC-AUC).
    \item We highlight a practical failure mode that can arise in QTN-based orbital classification. In our implementation, standard unit-norm scaling effectively removes the semi-latus rectum ($p$) information from quaternion sets. We propose the QTN$_p$ input vector to restore this physical dimension, which, for our models, appears to be an important condition for using quaternions in supervised learning.
    \item In our experiments, moving from an MEE-only representation to a triple-set MEE + PNC + QTN representation improves ROC-AUC from 0.789 to 0.858 and increases weighted F1 from 0.74 to 0.84. 
\end{itemize}


The remainder of this paper is structured as follows:
Section \ref{sec:methodology} will lay out the methodology of the experiment, and
highlight how this approach differs to those found in literature. Contained
within are subsections dedicated to each individual module of the computational
pipeline, along with a high level overview to illustrate how all the components
are orchestrated. Section \ref{sec:resultsanddiscussion} will report the
performance of the classification models and provide analytical insight into the
results. Finally, Section \ref{sec:conclusion} will summarize the key findings of
the experiment and suggest potential avenues for future work.

%% file: sections/method.tex
\section{METHODOLOGY}
\label{sec:methodology}




\subsection{Method Design and Computation}
\label{sec:computation}
\input{sections/computation.tex}

\subsection{Data Collection}
\label{sec:datacollection}
\input{sections/datacollection.tex}

\subsection{Breakup Model}
\label{sec:breakupmodel}

\input{sections/breakupmodel.tex}

\subsection{Fragment Filtering}
\label{sec:fragmentfiltering}
\input{sections/fragmentfiltering.tex}

\subsection{Fragment Propagation}
\label{sec:fragmentpropagation}
\input{sections/propagation.tex}
\subsection{Proper Element Extraction}
\label{sec:properelementextraction}
\input{sections/properelement.tex}
\subsection{Proper Element Filtering}
\label{sec:properelementfiltering}
\input{sections/properelementfiltering.tex}
\subsection{Supervised Classification with Neural Networks}
\label{sec:neuralnetworks}

\input{sections/neuralnetwork.tex}

%% file: sections/computation.tex
The methodology presented in \cite{Celletti2021} clearly establishes the 
potential of proper elements as suitable invariants for classification of 
orbital debris. The results showed a clear distinction between the fragment
families between two distinct fragmentation events, CZ-3 and Atlas V Centaur.
These events have been well documented in literature and the fragments have been
clearly well attributed to their respective parent objects. This prevents any
ambiguity and disputed labelling during classification tasks.
This is in contrast to the methodology in \cite{Wu2023} which utilizes two-line elements
(TLEs) from current SpaceTrack analyst databases. TLE datasets are known to
have some degree of inaccuracy in the data \cite{Bizalion2023,flohrer2008}, and
as with all empirical data collection, there is the potential for mislabelling.
However, the methodology in \cite{Wu2023} does build upon the foundation 
provided in \cite{Celletti2021} by scaling the classification task to include
several thousand fragments from an extensive catalog of known parent objects.


This work is inspired by the controlled and undisputed labelling of data
presented in \cite{Celletti2021} and the generalized scale of \cite{Wu2023}.
This work will aim to synthesise fragmentation data utilising parent objects
selected from real ephemerides, and simulating their breakup to create a
database of fragments with known parents. This enables a true evaluation of the
models used to classify fragments without attributing some known losses in
accuracy to data mislabelling in a large-scale simulation with several parent
objects.

\input{figures/pipeline.tex}

A general flow of the software components required in this
pipeline is illustrated in Figure \ref{fig:pipeline}.
Development of this work consisted primarily of integrating existing propagator
software into a highly parallel environment, implementing data
transformations, and the training of machine learning models. 
The majority of implementation was done with \texttt{Python} programming
language, with machine learning libraries such as \texttt{scikit-learn}
\cite{scikit-learn} and \texttt{PyTorch} \cite{pytorch}.
The UNSW High Performance Computing Cluster, Katana \cite{Katana} was used to
parallelise the propagation, data transformation, and proper element extraction
workloads. The training of neural networks was accelerated using a CUDA enabled
device.



%% file: figures/pipeline.tex
\begin{figure}[ht]
    \centering
    \includegraphics[width=0.48\textwidth]{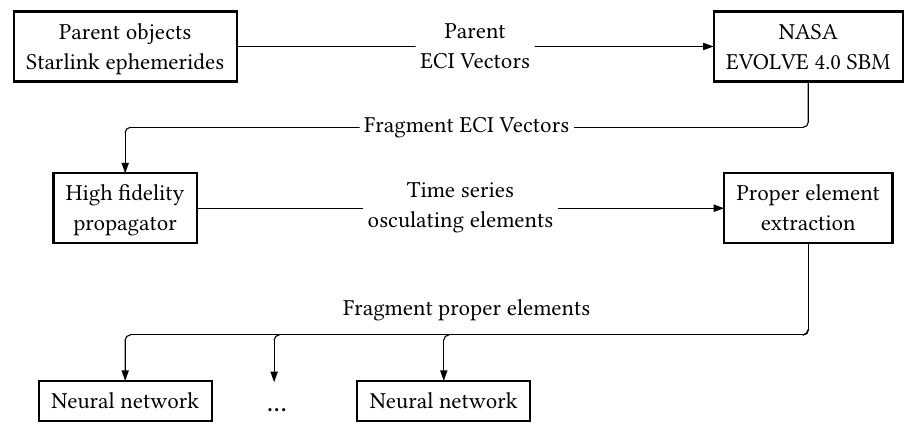}
    \caption{Compute pipeline.}
    \label{fig:pipeline}
\end{figure}

%% file: sections/datacollection.tex
A publicly available collection of active Starlink satellite
ephemerides was obtained from \hyperlink{www.space-track.org}{www.space-track.org} and transformed into Earth-centered inertial (ECI) state vectors. From this dataset, a random sample of 232 satellites was selected to serve as parent objects for fragmentation modeling. An example ECI vector for Starlink-4361 at Modified Julian Date (MJD) 60752.80256944 is shown in Table \ref{tab:starlinkparent}.
\input{tables/starlink_table.tex}
\input{tables/starlink_used.tex}

\input{figures/element_histograms.tex}
The selected satellites span a range of orbital parameters, as summarized in Table \ref{tab:starlink_used}. Figure \ref{fig:element_histograms} visualizes the distributions of semi-major axis, eccentricity, and inclination. The orbits are predominantly near-circular and distributed across distinct orbital planes. Most satellites occupy mid-inclination orbits, with a subset in highly inclined and sun-synchronous planes, and the majority reside in higher-altitude shells.

%% file: tables/starlink_table.tex
\begin{table}[ht]
    \centering
    \caption{Earth Centric Inertial Vector for Starlink-4361 at Modified Julian Date (MJD)
    60752.80256944}
    \label{tab:starlinkparent}
    \begin{tabular}{l r r r }
        \hline
        \centercell{\bf{Vector}} & \centercell{$x$} & \centercell{$y$} & \centercell{$z$} \\
        \hline
        $\vec{r}$ [km]   & 5657.634 & 1574.790 & -3707.461 \\
        $\vec{v}$ [km/s] & -3.529 & -2.214 & -6.327 \\
        \hline
    \end{tabular}
\end{table}
%

%% file: tables/starlink_used.tex
\begin{table}[ht]
    \centering
    \caption{Semi-major axis, eccentricity, and inclination ranges for the 
    Starlink satellites used in this work.}
    \label{tab:starlink_used}
    \begin{tabular}{lrrr}
    \hline
     & \centercell{$a$ [km]} & \centercell{$e$} & \centercell{$i$ [deg]} \\
    \hline
    max & 6955.281950 & 0.001818 & 97.798522 \\
    min & 6784.288052 & 0.000025 & 42.849187 \\
    \hline
    \end{tabular}
\end{table}

%% file: figures/element_histograms.tex
\begin{figure}[!h]
    \centering
    \includegraphics[width=0.48\textwidth]{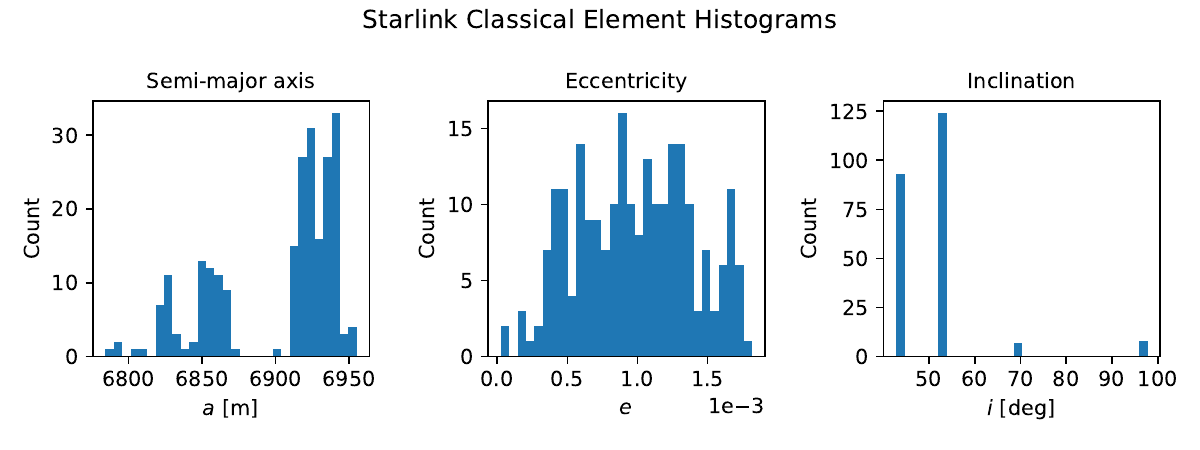}
    \caption{Starlink element histograms for semi-major axis, eccentricity, and
    inclinations. }
    \label{fig:element_histograms}
\end{figure}

%% file: sections/breakupmodel.tex
A wide range of works surrounding satellite breakup events refer and attempt to
improve upon the Standard Breakup Model (SBM) from the NASA EVOLVE 4.0 long term
satellite population model \cite{evolve}. It has been the subject for
improvement, with modifications to its parameters to better represent modern
satellites being made, or probabilistic models developed for improved space
traffic management \cite{Cimmino2021,Frey2021}. However, the original model
suffices for this experiment as it provides distinct fragments from which the
properties can be randomly sampled to create a database of simulated breakup
events.


The model specifies distinct variations of fragmentation distribution dependent
on the nature of the fragmentation event, namely whether the event occurred as a
result of collision or explosion, and parent body type, for example small
satellites or rocket bodies \cite{evolve}. This experiment attempts to
demonstrate model performance under a strictly controlled environment, thus only
the explosive events are considered.  For explosive events, the number of
fragments generated with a characteristic length less than $L_c^{min}$ is
described in Equation \ref{eqn:sbmnumber}:
\begin{align}
    N(L_c^{min}) = 6 S \left( L_c^{min}\right) ^{-k},
    \label{eqn:sbmnumber}
\end{align}
where $k=1.6$ for explosions, and $S=0.1$ for small satellites \cite{evolve}.

\begin{figure}[h]
    \centering
    \includegraphics[width=0.15\textwidth]{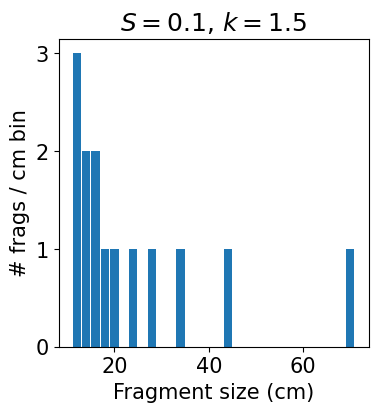}
    \includegraphics[width=0.16\textwidth]{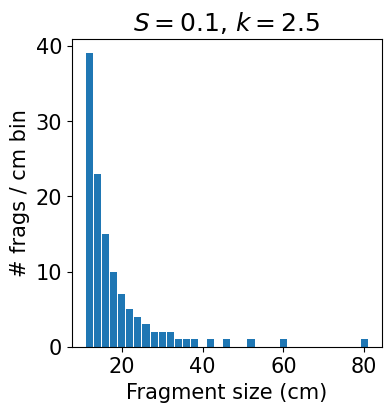}
    \includegraphics[width=0.16\textwidth]{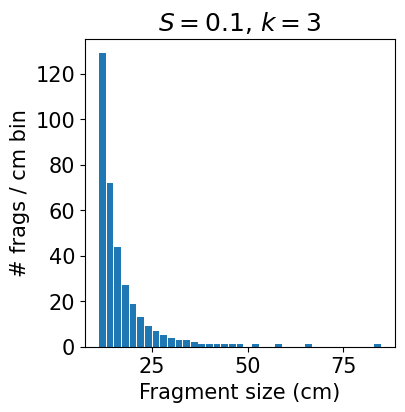}
    \caption{The granularity of fragmentation is shown by varying the $k$
    parameter. For explosions, Johnson suggests $k=1.6$ \cite{evolve}.}
    \label{fig:nfrags}
\end{figure}

To create a sample of all fragments generated with this model, we follow the
cumulative sampling methodology presented by Apetrii \cite{Apetrii2024} where
for a given $L_c^{min}$ in discrete [cm], i.e. the smallest fragment size
considered, we compute the fragment count for the 1 [cm] bin given by
$N(L_c^{min}) - N(L_c^{min} + 1)$. Continue for each [cm] bin, i.e. $N(L_c^{min}
+ i) - N(L_c^{min} + i + 1)$ until the aggregate sum of all bins equates to
$N(L_c^{min})$. This creates the fragment samples such as those seen in Figure
\ref{fig:nfrags}.

Introduce $\chi = \log_{10} (\frac{A}{M})$, where $\frac{A}{M}$ is the area-to-mass ratio. For $L_c > 11$ cm, $\chi$ follows a bimodal normal distribution $\mathcal{N}$, expressed in Equation \ref{eqn:sbmamr}:
\begin{align}
\chi \sim 
    \alpha \mathcal{N}( \mu_1, \sigma_1 ) + 
    (1 - \alpha) \mathcal{N}(\mu_2, \sigma_2).
\label{eqn:sbmamr}
\end{align}
Here, the sampling coefficient $\alpha$, distribution mean $\mu$, and standard
deviations $\sigma$ are piecewise functions of $\lambda_c = \log_{10} (L_c)$ and
parent body geometry, i.e., whether a launch vehicle upper stage, or a payload
spacecraft. A similar uni-modal normal distribution is used for fragments with
$L_c < 8$ cm, with a bridging function to patch the 8 to 11 cm discontinuity.
However, for this work, only fragments with $L_c > 11$ cm are considered.  All
distribution parameters are derived from \cite{evolve}.

Introduce $\nu = \log_{10}
(\Delta V)$, where $\Delta V$ is the explosive ejection velocity. $\nu$ follows a normal distribution, expressed in Equation \ref{eqn:sbmdeltav}:
\begin{align}
    \nu \sim 
        \mathcal{N} ( 0.2 \chi +1.85, 0.4 ).
\label{eqn:sbmdeltav}
\end{align}

To obtain a delta velocity vector (Equation \ref{eqn:deltaV} -
\ref{eqn:deltaV1}) from the magnitude $\Delta V$ found in Equation
\ref{eqn:sbmdeltav}, a spherical distribution was assumed as follows:
\begin{align}
    \label{eqn:deltaV}
    \Delta V_x &= \Delta V \sqrt{1 - c^2} \cos{\phi}, \\
    \Delta V_y &= \Delta V \sqrt{1 - c^2} \sin{\phi}, \\
    \Delta V_z &= \Delta V c,
    \label{eqn:deltaV1}
\end{align}
where $c \sim\text{Uniform}(0,1)$ and $\phi \sim \text{Uniform}(0, 2\pi)$.

After simulating fragmentation of all selected parent objects, the output is a
list of fragment data with each fragment entry containing its area-to-mass ratio, initial position and initial velocity. Parent and
resultant fragment examples are shown in Tables \ref{tab:starlinkparent} and
\ref{tab:starlinkfragment}. Table \ref{tab:starlinkfragment} captures the state
vector at the moment of the fragmentation event thus the fragment position
effectively remains identical to its parent in Table \ref{tab:starlinkparent}.
However, note the $\Delta V$ applied altering the velocity vector in comparison
to the parent object.

\input{tables/starlink_frag_table.tex}

%% file: tables/starlink_frag_table.tex
\begin{table}[!ht]
    \centering
    \caption{Earth Centric Inertial Vector for a fragment of Starlink-4361 at
    MJD 60752.80256944.}
    \label{tab:starlinkfragment}
    \begin{tabular}{l r r r}
        \hline
        \centercell{\bf{Vector}} & \centercell{{$x$}} & \centercell{$y$} & \centercell{$z$} \\
        \hline
        $\vec{r}$ [km]   & 5657.634 & 1574.790 & -3707.461 \\
        $\vec{v}$ [km/s] & -3.529 & -2.214 & -6.373 \\
        \hline
    \end{tabular}
\end{table}

%

%% file: sections/fragmentfiltering.tex
Due to the low orbital regime of the Starlink satellites, some fragments 
generated by the breakup model possessed insufficient total energy or a surface
area and thus drag coefficient too large to sustain an orbit during the full
duration of the propagation. A filtering method was devised to prevent excessive
computing resources from being wasted on re-entering objects. A sample of all
generated fragments was taken and propagated for the full experiment duration.
Re-entry status was determined at some cutoff for orbital height. Input
parameters for the fragment propagation were plotted for each fragment to derive
empirical thresholds for which most resultant fragments would likely possess
sufficient energy to maintain their orbit for the duration of the propagation.

The key parameters associated  were the magnitude of total velocity after
ejection and the fragment mass. The breakup model
provides equations required for the calculation of these parameters.
Theoretically, these parameters are not independent; however, in practice,
simple thresholds managed to select fragments in a satisfactory manner and
prevented wasting computing resources.

From this analysis, we derived thresholds of $V < 7.64 \text{ km/s}$ for velocity and $M > 5 \times 10^{-1} \text{ kg}$ for fragment mass. These thresholds are illustrated in Figure \ref{fig:threshold}. Although the fundamental mechanics behind
these thresholds are as of yet still unknown, a plausible cause could be
attributed to the effects of atmospheric drag. Given these objects populate the
LEO regime, the effects of atmospheric drag are significant.

\input{figures/threshold.tex}

%% file: figures/threshold.tex
\begin{figure}[h!]
    \centering
    \includegraphics[width=0.5\textwidth, trim={0 3cm 0 3cm}, clip]{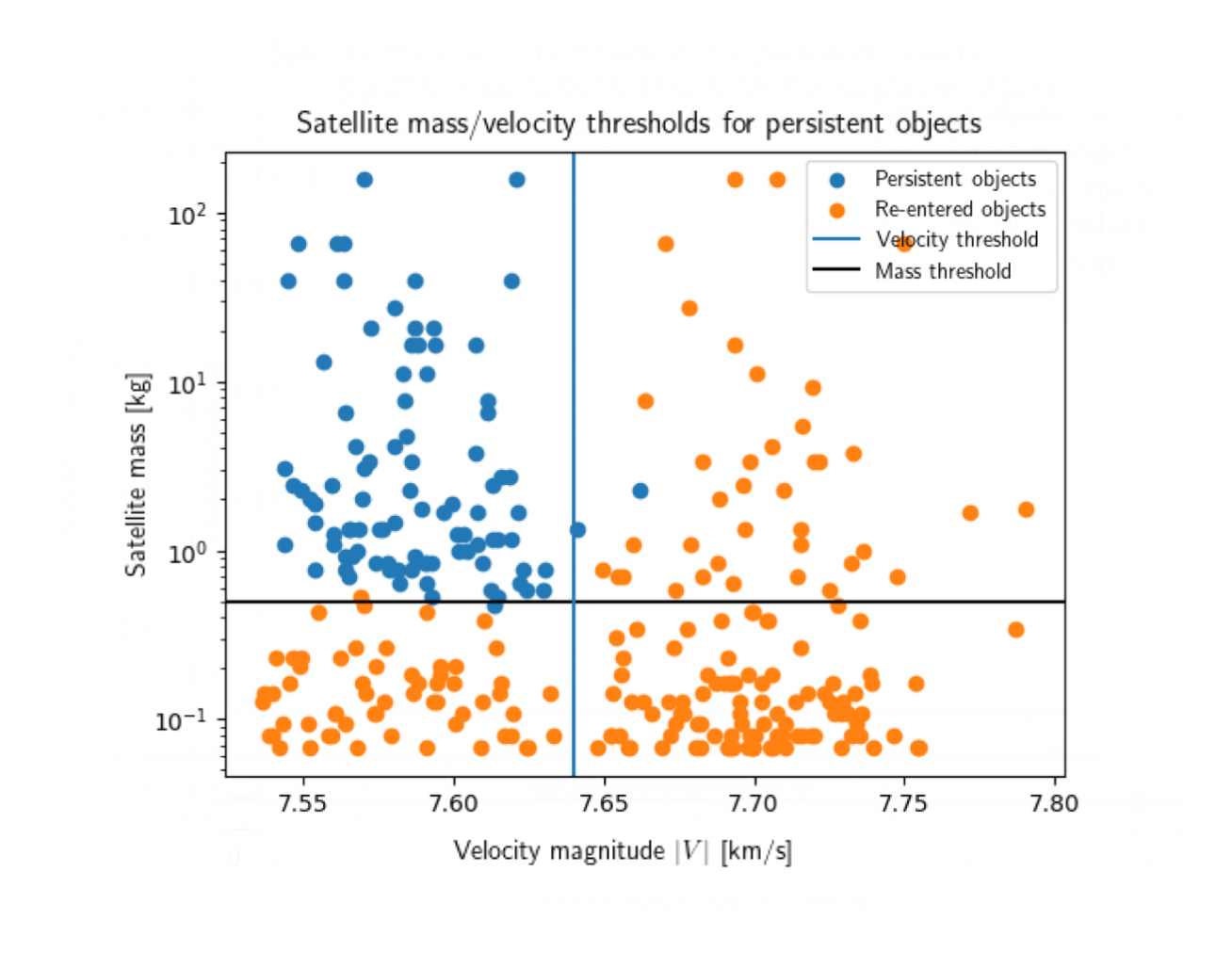}
    \caption{Scatter plot of a sample of propagated fragments, distinguished by
    re-entry status. Thresholds were empirically derived for the satellite mass
    and velocity.}
    \label{fig:threshold}
\end{figure}

%% file: sections/propagation.tex
Fragment propagation was performed with a high-fidelity orbit propagator
developed by Yang \cite{Yang2024}. The fragments ECI state vectors, see Table
\ref{tab:starlinkfragment}, were input and propagated at a 60 second time step
for 60 days under primary orbital perturbations, including a 100$\times$100
Earth gravitational harmonic model, third body lunisolar attractions,
atmospheric drag, solar radiation pressure, and Earth tidal effects \cite{orbitalmech2002}.  The output
given was a time series of ECI state vectors for each fragment.

%% file: sections/properelement.tex
The choice of coordinate system, i.e., orbital element set, fundamentally dictates the structure of the data presented to the learning algorithm and, consequently, its ability to discern patterns. 
After propagation, a list of ECI vectors were generated for each fragment over
the propagation period. These vectors were transformed into three sets of
orbital elements: MEE, PNC, and QTN \cite{Hintz2008}. Whereas previous RUO-classification studies computed proper elements only for MEE, we adapt the circle-fitting extraction approach to all three element sets, enabling a like-for-like comparison of their information content in a supervised-learning pipeline. Shown in Table \ref{tab:elementdef}, these alternate orbital elements are defined based on the classical orbital elements: semi-major axis $a$, eccentricity $e$, inclination $i$, right ascension of the ascending node $\Omega$, argument of perigee $\omega$, and mean anomaly $M$. $\mu$ is the Earth's gravitational parameter.
\input{tables/element_defs.tex}

The element sets were chosen because proper element extraction in LEO benefits from circle-fitting identities across several physically distinct representations (MEE, PNC, QTN), allowing us to test whether different dynamical fingerprints provide complementary information for classification.

For example, taking the
$(h,k)$ pair of elements from Table \ref{tab:elementdef}, we can form the
identity in Equation \ref{eqn:trigidentity}: 
\begin{align}
            e &= \sqrt{h^2 + k^2},
    \label{eqn:trigidentity}
\end{align}
which indicates the radius of the circle formed by the
$(h,k)$ pair represents a proper element of the eccentricity. Similarly, the $(p, q)$ pair
represent the proper inclination. The remaining pairs represent similar physical
quantities, though some more abstract than others, for example the action-related
quantities of PNC elements. It is worth noting that for the semi-major axis, and
its action-related counterpart in the PNC element set, that due to absence of 
short-period forcing frequencies, that the arithmetic mean of the time series is
sufficient to calculate its proper form \cite{Wu2023}. 
Table \ref{tab:elements} summarizes the specific element pairs that form these unit circle identities for all three element sets.
\input{tables/elements.tex}

The circle-fitting methods are taken from \cite{WuThesis} and
are shown in Algorithm \ref{alg:properelementinner} and Algorithm
\ref{alg:properelementouter}. In most cases, Algorithm
\ref{alg:properelementinner} had the tendency to isolate the short term
frequencies of the perturbed orbit as opposed to removing them.  This was
undesirable as proper element extraction is based on the isolation of the
fundamental forcing frequency of an orbit, not the short-period noise induced by
primary perturbations \cite{Gachet2017,Giacaglia1974}. Thus, the contiguous
window was substituted with interval sampling as described in Algorithm
\ref{alg:properelementinnerspaced} to increase the robustness of the inner loop.
This new interval sampling inner loop was used for the remainder of the
experiment. Visualizations of the contrasting inner loop algorithms are provided
in Figure \ref{fig:ptolemy}. 
\input{figures/ptolemy.tex}

\input{algorithms/innercontig.tex}
\input{algorithms/innerspaced.tex}
\input{algorithms/outer.tex}




%% file: tables/element_defs.tex
\begin{table*}[ht]
    \centering
    \caption{The three element sets chosen, and the elements calculated for the
    extraction of proper elements.}
    \label{tab:elementdef}
    \begin{tabular}{ l l l }

        \hline
        \centercell{\bf{MEE}} & \centercell{\bf{PNC}} & \centercell{\bf{QTN}} \\
        \hline
        $p=a(1-e^2)$                   & $\Lambda=\sqrt{\mu a}$                                          & $q_0=p^{1/4}\cos(i/2)\cos((\Omega + \omega + M)/2)$\\
        $h=e\cos(\Omega + \omega)$ & $\xi=e\sin(\Omega + \omega)(2\Lambda / (1+\sqrt{1-e^2}))^{1/2}$      & $q_1=p^{1/4}\sin(i/2)\cos((\Omega - \omega - M)/2)$\\
        $k=e\sin(\Omega + \omega)$ & $\eta=e\cos(\Omega + \omega)(2\Lambda / (1+\sqrt{1-e^2}))^{1/2}$     & $q_2=p^{1/4}\sin(i/2)\sin((\Omega - \omega - M)/2)$\\
        $f=i\cos(\Omega)$     & $u=\sin(i)\sin(\Omega)(2\Lambda\sqrt{1-e^2}/(1+\cos(i)))^{1/2}$ & $q_3=p^{1/4}\cos(i/2)\sin((\Omega + \omega + M)/2)$\\
        $g=i\sin(\Omega)$     & $v=\sin(i)\cos(\Omega)(2\Lambda\sqrt{1-e^2}/(1+\cos(i)))^{1/2}$ & $e_X=e\cos(M)$\\
        $\lambda = M + \Omega + \omega$        & $\lambda = M + \Omega + \omega$                                                  & $e_Y=-e\sin(M)$\\
        \hline

    \end{tabular}
\end{table*} 

%% file: tables/elements.tex
\begin{table}[ht]
    \centering
    \caption{Element sets selected contain pairs that allow formation of
    unit circle identities.}
    \label{tab:elements}
    \begin{tabular}{ c c c c }
        \hline

        \textbf{Element set} &
        \textbf{Pair 1} &
        \textbf{Pair 2} &
        \textbf{Pair 3 / Element} \\

        \hline

        MEE & $h, k$ & $f, g$ & $a$ \\
        PNC & $\xi, \eta$ & $u, v$ & $\Lambda$ \\
        QTN & $q_0, q_3$ & $q_1, q_2$ & $e_X, e_Y$ \\

        \hline
        
    \end{tabular}

\end{table}

%% file: figures/ptolemy.tex

\begin{figure*}[h!]
    \centering
    \includegraphics[width=0.25\textwidth]{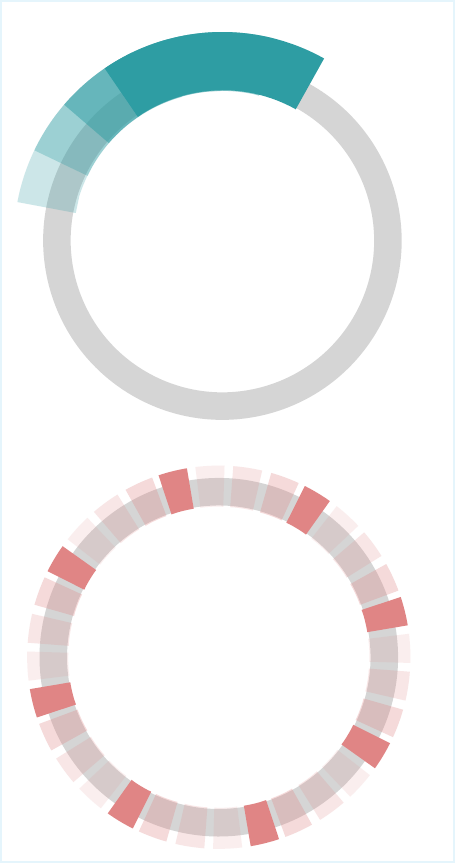}
    \includegraphics[width=0.38\textwidth, trim={6cm 15cm 5cm 2cm}, clip]{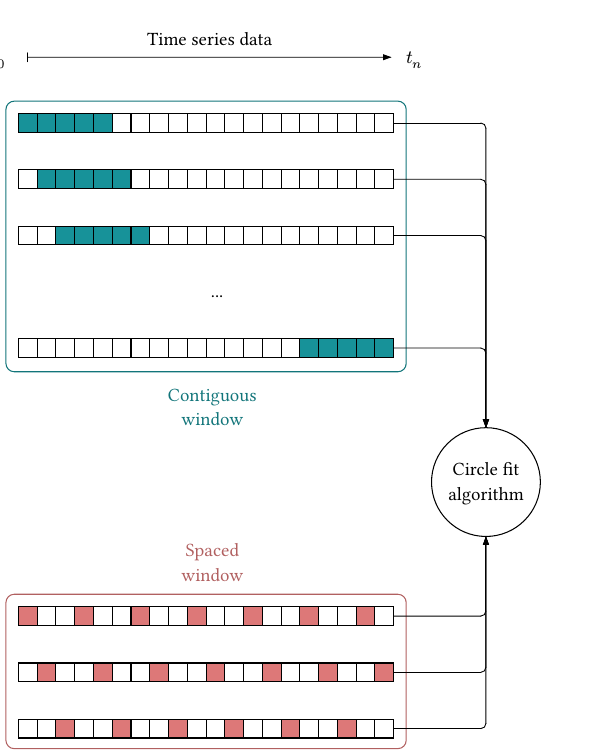}
    \caption{Diagrams for the contiguous and spaced window inner loop
    algorithms.}
    \label{fig:ptolemy}
\end{figure*}

%% file: algorithms/innercontig.tex
\begin{algorithm*}[ht]
    \caption{Proper element extraction inner loop, contiguous window}
    \label{alg:properelementinner}
    \begin{algorithmic}
        \State \textbf{Input:} Window size, Sequence of paired elements
        \State \textbf{Output:} Proper element, RMSE
        \State $seq \gets$ sequence of elements
        \State $L \gets length(seq)$
        \State $n \gets \text{window size}$
    
        \For{$i = 0, 1, ..., L-n$}
            \State $data \gets seq[i : i + n]$
            \Comment{Selecting a window from the sequence}
            \State $r_i,\space \sigma_i \gets circleFit(data)$
            \Comment{Circle fitting algorithm returns radius and error}
            \State $circleArray[i] \gets r_i,\space \sigma_i$
        \EndFor
        
        \State $r_n,\space \sigma_n \gets mean(circleArray)$
        \Comment{Find mean radius, find mean error}
    
        \Return $r_n,\space \sigma_n$
    \end{algorithmic}
\end{algorithm*}

%% file: algorithms/innerspaced.tex
\begin{algorithm*}[ht]
    \caption{Proper element extraction inner loop, interval window}
    \label{alg:properelementinnerspaced}
    \begin{algorithmic}
        \State \textbf{Input: } Window spacing, sequence of paired elements
        \State \textbf{Output: } Proper element, RMSE
        \State $seq \gets$ sequence of elements
        \State $S \gets$ window spacing
        
        \For{$i = 0, 1, ..., S$}
            \State $data \gets seq[i::S]$
            \Comment{Selecting every $S$-th epoch starting at index $i$}
            \State $r_i,\space \sigma_i \gets circleFit(data)$
            \Comment{Circle fitting algorithm returns radius and error}
            \State $circleArray[i] \gets r_i,\space \sigma_i$
        \EndFor

        \State $r_n,\space \sigma_n \gets mean(circleArray)$
        \Comment{Find mean radius, find mean error}
        
        \Return $r_n,\space \sigma_n$
    \end{algorithmic}
\end{algorithm*}

%% file: algorithms/outer.tex
\begin{algorithm*}[!h]
    \caption{Proper element extraction outer loop}
    \label{alg:properelementouter}
    
    \begin{algorithmic}
        \State \textbf{Input:} List of window lengths, Sequence of trigonometrically paired orbital elements
        \State \textbf{Output:} Highest confidence proper element
        \State $winLens \gets$ List of window lengths
        \State $numWins \gets length(winLens)$
        \State $seq \gets$ sequence of elements
        \For{$i = 0,1,2,...,numWins $}
            \State $rList[i], \sigma List[i] \gets innerLoop(seq, winLens[i])$
        \EndFor
        \State $minIndex = index(min(\sigma List))$
        \State \textbf{return} $rList[minIndex]$
    \end{algorithmic}
\end{algorithm*}

%% file: sections/properelementfiltering.tex
After determining the circles of best fit, the results were graphically
assessed. Another filtering step was devised to ensure that the proper element
data used in later stages was clean and was a physically accurate representation
of the dynamics of the fragments.  It is worth noting, PrattSVD \cite{Pratt1987} was the
implemented circle fitting algorithm. PrattSVD is an algebraic circle fit
algorithm and was chosen for its low computational cost. Although the stability
of a geometric fit, such as least-squares, is superior to algebraic fits, the
computational cost was a limiting factor, and with the filtering process
described in the following section, geometrically inconsistent fits could be
excluded. Examples of the inconsistent algebraic fits can be seen in Figure
\ref{fig:filterproperelements}. 

The filtering algorithm discriminates circle fits by checking two properties;
whether a random subset of the time series data actually lie within a given
bandwidth surrounding the circle, and whether those points have sufficient
radial coverage. Radial coverage is determined by dividing radial space into bins
and checking for sufficient number of occupied bins. The results of filtering a
sample of the dataset on the $h, k$ proper element are illustrated in Figure
\ref{fig:filterproperelements}. The blue track represents the epicyclic
osculating element time-series data. Superimposed is the best circle of best fit
as found by applying Algorithm \ref{alg:properelementouter} to this data. A red
highlight indicates a satisfactory fit.

\input{figures/filterproperelements.tex}

%% file: figures/filterproperelements.tex
\begin{figure*}
    \centering
    \includegraphics[width=0.9\textwidth]{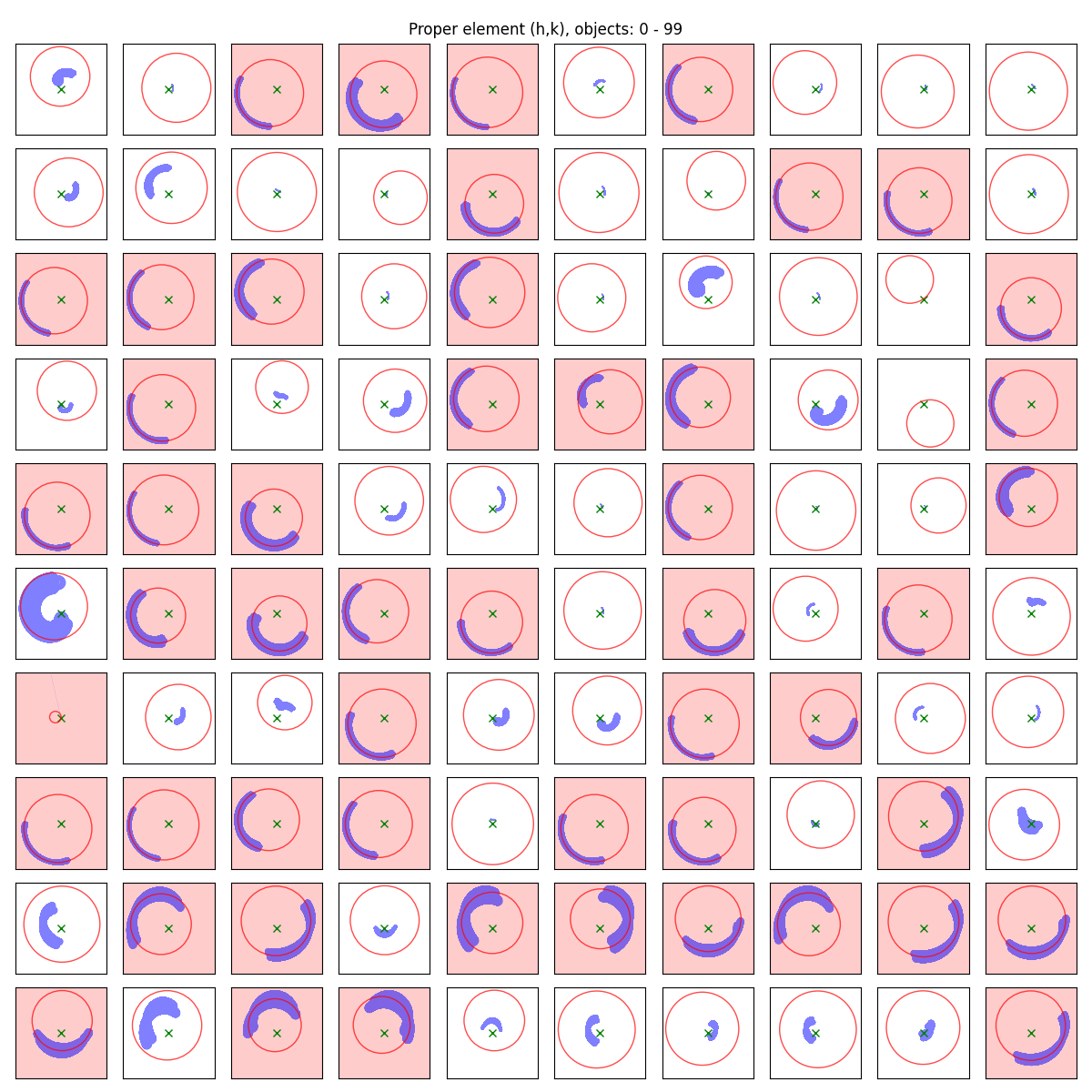}
    \caption{Results of filtering MEE proper elements $h, k$. Red fills indicate
    a pass for that fragment. The blue smear is the time series of the elements.
    }
    \label{fig:filterproperelements}
\end{figure*}

%% file: sections/neuralnetwork.tex

In a Cartesian space such as ECI vectors, Euclidean distance is the intuitive choice for proximity. In classical orbital-element space, however, “distance” is not uniquely defined. Following convention in asteroid/debris taxonomy, the Zappal\`a-style threshold function \cite{Zappala1990}, shown in Equation \ref{eqn:distancefunction}, is used to measure separation between proper elements and to provide a baseline for taxonomy:
\begin{align}
    d_Z = n_\text{ave} a_\text{ave} \sqrt{ k_1 \left(\frac{\delta a}{a_\text{ave}}\right) + k_2 (\delta e)^2 + k_3 (\delta i) ^ 2 }.
    \label{eqn:distancefunction}
\end{align}
Here, $\delta(\cdot)$ denotes pairwise differences of proper elements; $a_{\text{ave}}$ is the mean semi-major axis and $n_{\text{ave}}=\sqrt{\mu/a_{\text{ave}}^{3}}$ the mean motion. Coefficients $k_1,k_2,k_3$ weight each element's contribution (e.g., $k_1=5/4$, $k_2=2$, $k_3=2$).

In adherence to \cite{Zappala1990}, the original multilayer perceptron feedforward neural network (MLPFNN) in \cite{Wu2023} consumes first- and second-order differences of the proper elements from fragments $(a_1,e_1,i_1)$ and $(a_2,e_2,i_2)$, as shown in Equation \ref{eqn:wuinput}, to surrogate the classical distance metric. The objective of that MLPFNN was to emulate the discriminatory power of Equation \ref{eqn:distancefunction} while circumventing the empirical $k_{1-3}$ parameter tuning, in favor of the context-aware, nonlinear function of $\delta$ terms provided by the MLPFNN. This choice ensured the classifications were consistent with the classical literature while leveraging the capabilities of neural networks.
\begin{align}
    \text{Input} = (\delta a, \delta e, \delta i, (\delta a)^2, (\delta e)^2, (\delta i)^2)
    \label{eqn:wuinput}
\end{align}

Following \cite{Wu2023}, we adopted the MLPFNN as a starting point: an input layer of width $2\times(\text{\# elements})$, $L$ hidden layers of width $N$ with \textit{ReLU} activations, and a two-unit softmax output giving the probabilities of Classes 0/1 (different/same parent). Models are trained with cross-entropy and \textit{adamW}, with the learning rate $r$ and batch size $B$ treated as explicit hyperparameters. The generic architecture is shown in Figure \ref{fig:nntopology}, and the full hyperparameter search spaces are summarized in Table \ref{tab:hyperparameterspace}.
\input{figures/nn.tex}

However, two issues motivate a different input design for LEO debris. (i) {Redundancy:} single-hidden-layer MLPFNNs are universal approximators \cite{Hornik1989}; explicit quadratic terms are unnecessary and can amplify noise. (ii) {State completeness:} pure differences discard absolute location in proper-element space. In crowded LEO, absolute context helps separate neighbouring families; discarding it can reduce recall for minority families and degrade physical interpretability \cite{Raissi2019}. Accordingly, we feed the network the paired absolute proper elements for both fragments (preserving the full state; see the following Equations \ref{eqn:MEE}--\ref{eqn:QTN_p}). The learner then internalizes linear/nonlinear relations rather than relying on hand-crafted squares\,—that is, explicit quadratic terms chosen \textit{a priori} such as $(\delta e)^2$ and $(\delta i)^2$ in Equation \ref{eqn:distancefunction} and the $(\delta\cdot)^2$ features in Equation \ref{eqn:wuinput}. Rather than fixing a global quadratic geometry of separation, the network learns the effective interactions from data, which can be quadratic or higher order and can vary across regions of proper element space.

We trained separate models on single, double, and triple element-set combinations (Equations \ref{eqn:MEE}--\ref{eqn:QTN_p}). Architecture and hyperparameters are tuned with \texttt{ray-tune} library \cite{liaw2018tune} over the search space in Table~\ref{tab:hyperparameterspace}, selecting the checkpoint with the best validation F1-score given by Equation \ref{eqn:f1score} and using early stopping to prevent overfitting.
\input{tables/hyperparameterspace.tex}
Since data input was required in pairwise form, a pairwise list was generated
for all possible fragment pairs and labelled with Class 1 ($n_1$) if the two
fragments in the pair are from the same parent, and Class 0 ($n_0$) otherwise.

Due to overwhelming variation in order of magnitude of orbital elements, 
 preprocessing via data normalization is required to prevent numerically dominant 
 elements from skewing the neural network training. It is standard practice to 
 scale data such that the sample mean is located at 0 and the sample standard
 deviation is 1. This preprocessing technique is applied to all individual
 elements, except for angular elements and the quaternion vector. To scale these
 quantities, their physical representation must be considered. For example,
 values near 0 or $2\pi$ are physically adjacent, but on a linear space, the
 neural network is only informed that 0 and $2\pi$ are distant. Scaling for
 angular elements includes $\sin$/$\cos$ encoding to remove discontinuities in
 the data at 0 and $2\pi$. The quaternion vector is scaled to ensure its
 $\text{norm}_2 = 1$.

As defined in Table~\ref{tab:elementdef}, the semi-latus rectum $p$ is encoded within $(q_0,q_1,q_2,q_3)$. 
However, neural networks require input data to be normalized (typically $\mu=0$, $\sigma=1$ or unit norm) to prevent gradient explosion and ensure convergence. Standard normalization of the quaternion vector forces it to unit length, effectively deleting the variable $p$ and collapsing all orbits to a single ``size". This renders the feature set physically incomplete for distinguishing families based on orbital scale. To mitigate this, $p$ was appended to form the {QTN$_p$} element set, see Equation \ref{eqn:QTN_p}. To assess redundancy of explicit quadratic terms, {MEE$^2$} was also constructed by adding squared components to MEE, see Equation \ref{eqn:MEE+MEE2}. We then assembled datasets with single, double, and triple combinations, summarized in Equations \ref{eqn:MEE}--\ref{eqn:QTN_p}. Here $\p(\cdot)$ denotes the proper element of the specified quantity for each fragment (subscripts 1 and 2). This experimental design explicitly isolates the effect of (i) adding new proper-element sets beyond MEE (PNC and QTN), (ii) restoring state completeness ({QTN$_p$}), and (iii) adding explicit quadratic terms ({MEE$^2$}), so that we can quantify how each choice influences supervised classification quality.

\subsubsection{Single element sets:}
\begin{align}
    	\text{MEE} =& (\p(a)_1, \p(h,k)_1, \p(f,g)_1, \notag \\
        &\p(a)_2, \p(h,k)_2, \p(f, g)_2),
    \label{eqn:MEE}
\end{align}

\begin{align}
    \text{PNC} =& (\p(\Lambda)_1, \p(\xi,\eta)_1, \p(u,v)_1, \notag \\
    &\p(\Lambda)_2, \p(\xi,\eta)_2, \p(u,v)_2),
    \label{eqn:PNC}
\end{align}
\begin{align}
    \text{QTN} =& (\p(q_0,q_3)_1, \p(q_1,q_2)_1, \p(e_X,e_Y)_1, \notag \\
     &\p(q_0,q_3)_2, \p(q_1,q_2)_2, \p(e_X,e_Y)_2).
    \label{eqn:QTN}
\end{align}

\subsubsection{Double element sets:}
\begin{align}
    \text{MEE + PNC} = (\text{MEE}, \text{PNC}),
    \label{eqn:MEE+PNC}
\end{align}
\begin{align}
    \text{MEE + QTN} = (\text{MEE}, \text{QTN}),
    \label{eqn:MEE+QTN}
\end{align}
\begin{align}
    \text{PNC + QTN} = (\text{PNC}, \text{QTN}).
    \label{eqn:PNC+QTN}
\end{align}

\subsubsection{Triple element set, and others:}
\begin{align}
    \text{MEE + PNC + QTN} = (\text{MEE}, \text{PNC}, \text{QTN}),
    \label{eqn:MEE+PNC+QTN}
\end{align}
\begin{align}
    \text{MEE + MEE}^2 =& (\text{MEE}, \p(a)_1^2, \p(h,k)_1^2, \p(f,g)_1^2, \notag \\
    &\p(a)_2^2, \p(h,k)_2^2, \p(f, g)_2^2),
    \label{eqn:MEE+MEE2}
\end{align}
\begin{align}
    \text{QTN}_p = (\text{QTN}, p).
    \label{eqn:QTN_p}
\end{align}

The dataset used for all training, validation, and testing was identical for all models, and was constructed from a split of original dataset based on parents' identity.  The training set is randomly resampled to balance positive and negative pairs, avoiding bias toward the majority Class 0 ($n_0$). Validation and test sets preserve the natural class imbalance to reflect operational conditions. See Figure \ref{fig:dataset} for the entire dataset
construction. 
\input{figures/dataset.tex}



%% file: figures/nn.tex
\begin{figure}[ht]
    \centering
    \includegraphics[width=0.45\textwidth]{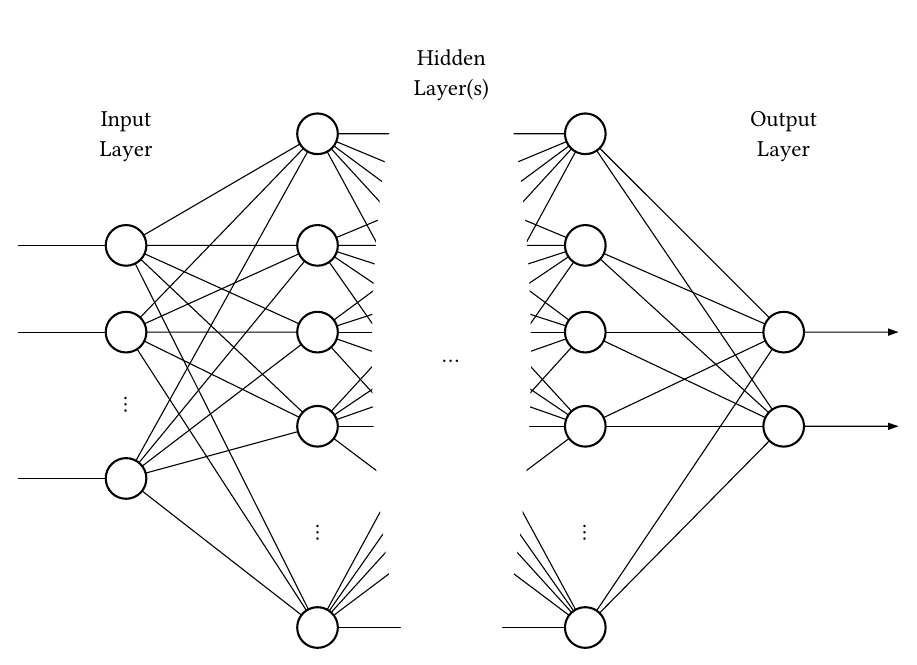}
    \caption{Standard Multilayer perceptron feedforward neural network architecture.}
    \label{fig:nntopology}
\end{figure}

%% file: tables/hyperparameterspace.tex
\begin{table*}[ht]
    \centering
    \caption{Hyperparameter search space}
    \label{tab:hyperparameterspace}
    \begin{tabular}{lcc}
        \hline
        
        \textbf{Hyperparameter} &
        \textbf{Symbol} &
        \textbf{Search Space / Distribution} \\

        \hline

        Learning Rate & $r$ & $\text{LogUniform}(10^{-4}, 10^{-1})$ \\
        Batch Size & $B$ & $\{512 k \mid k \in \mathbb{Z}^+, 1 \le k \le 8\}$ \\
        Layer Width & $N$ & $\{128 k \mid k \in \mathbb{Z}^+, 1 \le k \le 8\}$ \\
        Number of Layers & $L$ & $\{k \mid k \in \mathbb{Z}^+, 1 \le k \le 8\}$ \\
        
        \hline
    \end{tabular}
\end{table*}

%% file: figures/dataset.tex
\begin{figure}[ht]
    \centering
    \includegraphics[width=0.48\textwidth]{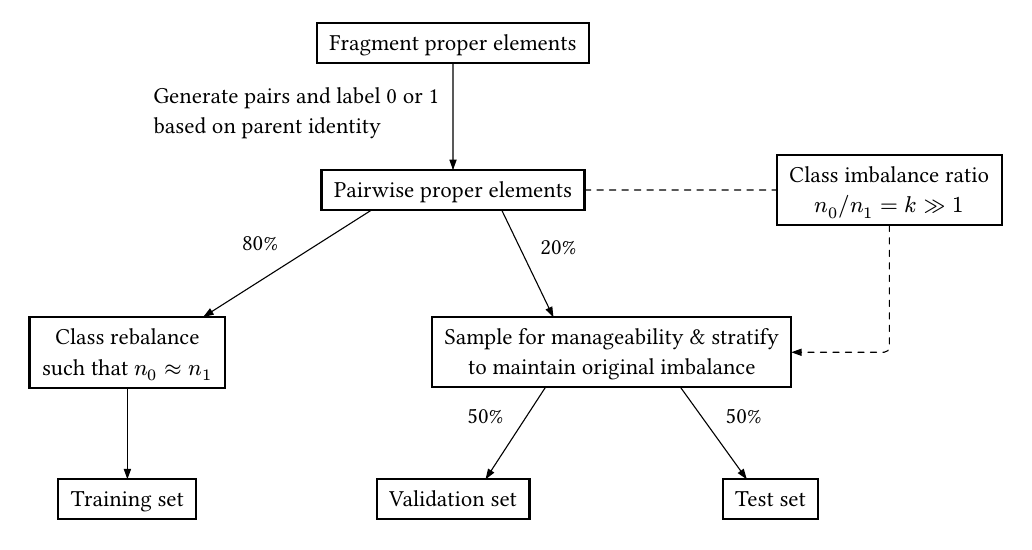}
    \caption{Dataset construction.}
    \label{fig:dataset}
\end{figure}

%% file: sections/results_and_discussion.tex
\section{Results and Discussion}
\label{sec:resultsanddiscussion}
\subsection{Evaluation Metrics}
The neural classifiers MLPFNN output a probability \(p_b=\Pr(\text{same
parent}\mid\text{features})\) for each fragment pair. Applying a decision
threshold \(\tau\) yields a binary label: pairs with \(p_b\ge\tau\) are
predicted as same-parent, otherwise not. We compare models using accuracy,
precision, recall, and \(F_1\) as defined in
Equations. \ref{eqn:accuracy}--\ref{eqn:f1score}:
\begin{equation}
    \label{eqn:accuracy}
    \text{Accuracy} = \dfrac{TP + TN}{TP + TN + FP + FN},
\end{equation}
\begin{equation}
    \label{eqn:precision}
    \precision = \dfrac{TP}{TP + FP},
\end{equation}
\begin{equation}
    \label{eqn:recall}
    \recall = \dfrac{TP}{TP + FN},
\end{equation}
\begin{equation}
    \label{eqn:f1score}
    F_1 = \dfrac{2 \times \precision \times \recall}{\precision + \recall},
\end{equation}
where $TP$, $TN$, $FP$, $FN$ denote true positives
(TP), true negatives (TN), false positives (FP), and false negatives (FN),
respectively. Accuracy is the proportion of correct predictions for both
positive and negative classes over all cases. However, under severe class
imbalance it can be misleading. Precision is the models accuracy within a single
class. Recall measures the proportion of TPs the model was able to capture from
all positive cases. The $F_1$ score is the harmonic mean of precision and recall
computed per class. Because the data are imbalanced, we report both macro and
weighted $F_1$:
\begin{align}
F_1^{\text{macro}} &= \tfrac{1}{2}\Big(F_1^{(+)} + F_1^{(-)}\Big), \label{eqn:f1macro}\\
F_1^{\text{weighted}} &= \frac{n_{+}\,F_1^{(+)} + n_{-}\,F_1^{(-)}}{n_{+} + n_{-}}, \label{eqn:f1weighted}
\end{align}
where $F_1^{(+)}$ and $F_1^{(-)}$ are the per-class $F_1$ for the positive
(“same parent”) and negative classes, and $n_{+},n_{-}$ are their supports
(instance counts).  Macro $F_1$ reflects balanced attribution capability
irrespective of prevalence; weighted $F_1$ reflects expected deployment
performance under the observed class distribution. A large gap (weighted $\gg$
macro) indicates the model still underperforms on the rare positive class, even
if overall performance appears high.
\input{tables/classificationreport.tex}
\input{figures/cmat.tex}
\subsection{Results}
Accuracy, macro $F_1$, and weighted $F_1$ results with a baseline threshold of
$\tau=0.5$ on $p_b$ are summarized in Table \ref{tab:classificationreport}.
These results also quantify the impact of enlarging the proper-element
representation from the single-set MEE proper space used in previous neural
approaches to multi-set combinations that include PNC and QTN. The MEE + PNC +
QTN model attains the highest accuracy and both F1 scores, indicating
complementary physical information across element sets: for example, at
$\tau=0.5$ on $p_b$ it improves accuracy from 0.61 (MEE) to 0.75, macro F1 from
0.42 to 0.49, and weighted F1 from 0.44 to 0.84, while also increasing area
under the receiver operating characteristic curve (ROC-AUC) from 0.789 to 0.858.
Single-set models (MEE, PNC) are moderate, while {QTN} alone underperforms.
Appending $p$ to form {QTN$_p$} restores state completeness and markedly
improves both macro and weighted $F_1$, bringing it closer to the other
single-set models. {MEE+PNC} and {MEE+QTN} are strong and close to the
triple-set model, consistent with complementary content and the observed
diminishing returns. {MEE$^2$} offers only marginal change over MEE, consistent
with the expectation that the network learns non-linearities without
hand-crafted squares. Across models, weighted $F_1$ substantially exceeds macro
$F_1$ ($\Delta \approx 0.2 - 0.35$), indicating strong performance on the
majority class but remaining difficulty on the rare positive class; this
motivates threshold tuning ($\tau = \tau^*$ for $F_1$) to improve positive-class
precision/recall. For instance, shifting the decision threshold to the
$F_1$-optimal value of $\tau=0.99$ for the MEE + PNC + QTN model reduces the FP
rate from roughly $25.52\%$ (at $\tau=0.5$) to approximately $9.53\%$. This
reduction significantly lowers the rate of false alarms while maintaining a
recall of roughly $50.52\%$.

Figure \ref{fig:cmat} illustrates the confusion matrices for the classifications
made by each model at different thresholds. Adjusting the threshold for
different values leads to changes in the ratios between TN (top-left), TP
(bottom-right), FP (top-right), and FN (bottom-left). The left column fixes a
baseline threshold of $\tau=0.5$ on $p_b$. In this imbalanced pairwise setting,
$\tau=0.5$ yields relatively higher recall but lower precision, visible in the
upper right quadrant across models. The right column uses a macro-$F_1$-optimal
threshold $\tau^* = \arg\max_{\tau} F_1^{\text{macro}}$ selected on the
validation set by sweeping $\tau$ over $\{0.01,0.02,\dots,0.99\}$. This
operating point explicitly maximises macro $F_1$ under class imbalance, trading
some TNs for additional TPs, and reduces over-/under-linking behaviour without
changing the overall model ranking.

In addition to confusion matrices, the area under the receiver operating
characteristic curves (ROC) can be used to examine the performance of models in
against a random classifier, which would be making classification by random
selection. The greater the area under the curve (AUC) the more performant the
model is at binary classification as opposed to random guessing. As ROC-AUC
approaches unit area, the performance against the random classifier increases.
The ROC-AUCs for all models can be found in Figure \ref{fig:roc}. The best
performing model is the MEE + PNC + QTN model. In general, models with access to
more element sets perform better than those with less. The AUC for the MEE model
with both first and second-order terms provides only minor benefit to the
classification performance, negligible compared to the benefit of adding another
set of proper elements. The QTN model lacking the semi-latus rectum $p$
information performs significantly worse than the QTN model with access to that
information.

\input{figures/roc.tex}

From the confusion matrices in Figure \ref{fig:cmat}, the ROC curves in Figure
\ref{fig:roc}, and the metrics in Table \ref{tab:classificationreport}, the
consensus is that the highest scoring model is the MEE + PNC + QTN model. In
general, the models with access to more elements sets make better predictions.
This is evidence that proper elements of different element sets with different
physical representations provide independent information for a neural network to
better classify input data.

Another observation is that the QTN$_p$ set outperforms the original QTN set,
while the set with second order MEE$^2$ terms boasts only a slight increase to
performance compared to the original MEE set, and in general only performs as
well as single-set models. However, the extent of the benefit gained by adding
more element sets appears diminishing, with the increase in performance from the
triple-element-set model to double-element-set model being minimal in comparison
to the benefit from going from single-set to double-set.

Taken together, Table \ref{tab:classificationreport} and Figure \ref{fig:roc}
suggest the following provisional design guidance for RUO classifiers trained in similar LEO-like, synthetic settings: (i) in our experiments, using more than one proper-element set is consistently better
than any single-set model; (ii) restoring state completeness (e.g. QTN$_p$ vs
QTN) can matter as much as adding a second element set; and (iii) explicit
quadratic terms (MEE$^2$) provide only marginal improvements relative to the
gains from expanding the proper-element set.

\subsection{Discussion}

\subsubsection{Domain shift sensitivity}
A key challenge is the potential domain shift between synthetic training data
and real-world breakups. Our training relies on the EVOLVE 4.0 model for pure
explosive fragmentation events and a specific granularity, which uses empirical
parameters derived from historical data. Modern satellite materials and varying
explosive energies may yield fragment distributions that differ from these fixed
predictions. If real-world debris clouds deviate significantly, model
performance may degrade.  Additionally, the fragments synthesized for training
originate from a small number of LEO orbital shells, seen in Table
\ref{tab:starlink_used} and Figure \ref{fig:element_histograms}. The limited
breadth of this dataset hinders the generalizability of this model to other
orbital regimes, for example the particularly vital geostationary regime.
Future work should aim to ensure representation from varying terrestrial
orbital regimes, and by introducing a dynamic system to capture the evolution of
satellites and their fragmentation (e.g., scaling factor $S$ and distribution
shape $k$). These contributions will improve how robust these models are when 
faced with domain shift.

\subsubsection{Input representation analysis}
In previous work, Wu \cite{Wu2023} included higher-order terms in the input to
the neural network. In the literature, significant emphasis is placed on the
thresholding function expressed in Equation \ref{eqn:distancefunction}, proposed
by Zappal\`a et al. \cite{Zappala1990}. That study adapts this distance
function in a neural-network-based approach, using first- and second-order
differences of the proper elements. However, by including only difference terms,
the amount of information available to the neural network is reduced.
Furthermore, a neural network is capable of modelling simple linear interactions
between its inputs, thus including raw proper elements for each fragment in the
pair preserves information without obfuscating physical relation between the
fragments.  Additionally, it was demonstrated that neural networks are capable
of modelling second-order relations. Figure \ref{fig:roc} and the metrics in
Table \ref{tab:classificationreport} illustrate the negligible increase in
performance by incorporating second-order terms into the model.  Loss of state
completeness is shown to cause detriment to model performance as highlighted by
the reduced accuracy from 0.59 to 0.31 shown in Table
\ref{tab:classificationreport} and reduced ROC-AUC from 0.77 to 0.69 shown in
Figure \ref{fig:roc}. The restoration of performance in the QTN$p$ model
confirms that neural networks cannot recover latent physical variables that have
been mathematically removed by preprocessing. 

\subsubsection{Multi-set complementarity}
To our knowledge, this is the first study to systematically compare multiple
proper-element sets, including PNC and QTN, in a supervised-learning framework
for debris-family classification. Previous work validated the feasibility of MEE
proper elements for RUO classification, but did not examine whether alternative
or combined proper-element spaces could offer higher discriminative power or
robustness as the circumterrestrial environment evolves. Our findings show that
multi-set representations do indeed yield measurable gains over the best
MEE-only baseline and that these gains are attributable to genuinely
complementary dynamical information, rather than to ad-hoc higher-order
features.


\subsubsection{Operational considerations} From a computational perspective,
utilizing the triple-set model (MEE + PNC + QTN) approximately triples the
runtime of the proper element extraction phase compared to the baseline MEE-only
model. However, this overhead is negligible in the context of the full pipeline.
The dominant computational bottleneck remains the high-fidelity numerical
propagation of fragment state vectors over 60 days. The subsequent
transformation of these vectors into element sets relies on analytical algebraic
identities (Table \ref{tab:elementdef}), which are computationally instantaneous
relative to the integration steps.

It is also worth mentioning that the interpretation of the confusion matrices in
Figure \ref{fig:cmat} can be variable depending on the context of the task it is
applied to. Here, the optimal thresholds are chosen purely for maximizing $F_1$
score, however different tasks necessitate choosing different metrics. A
cost-benefit analysis of minimizing false-positives as opposed to
false-negatives would reveal the optimal classification thresholds to be used 
when using these models. For example, an exploratory task could utilize the
model as a first pass to reduce the amount of manual investigation into
fragments, and would seek to minimize false-negatives to the cost of discovery
outweighing that of a larger search space.

\subsubsection{Benchmarking and generalizability}
Comparison with other similar approaches is difficult due to the limited
attempts in literature. Wu \cite{Wu2023} reports a 90\% accuracy for the model
trained on real-world analyst objects. However, without other metrics for the
severely imbalanced dataset in question, comparison is qualitative at best.
Nonetheless, the work in this paper aligns with the suggestions made for the
approach in \cite{Wu2023}, where increasing the element set and dimensionality
of the data available to the model could lead to an increase in classification
performance.

Only by demonstrating robust predictions in varying environments would a model
be considered suitable for general application, and with differences in
performance being illustrated, it must be stated that despite its capability in
complex thresholding for RSO classification in the context the present
environment, the current approach to training neural networks is still
vulnerable to underperforming.  A truly general model would have to encompass
not only the distribution and dynamics of the current environment, but also
predict the evolution of those properties. As stated in Section \ref{sec:intro},
the demand for the capability to classify RUOs is increasing, and it may occur
where the known distribution of RSOs is dissimilar to that of RUOs. As such, to
achieve the goal of creating a model capable of such a task will require a
multidisciplinary collaboration with breakup, collision probability, and space
traffic modelling.



%% file: tables/classificationreport.tex
\begin{table*}[ht]
    \centering
    \caption{Classification report summary for all models at a threshold of $\tau = 0.5$.}
    \label{tab:classificationreport}
    \begin{tabular}{ l r r r r}
        \hline

    	\textbf{Model} &
    	\textbf{Accuracy} &
    \begin{tabular}[r]{@{}r@{}}\textbf{Macro} \\ \textbf{Averaged} $\mathbf{F_1}$\end{tabular} &
    \begin{tabular}[r]{@{}r@{}}\textbf{Weighted} \\ \textbf{Averaged} $\mathbf{F_1}$\end{tabular} \\
        
        \hline

        MEE & 0.61 & 0.42 & 0.74 \\
        PNC & 0.59 & 0.40 & 0.72 \\
        QTN & 0.31 & 0.26 & 0.45 \\

        MEE + PNC & 0.72 & 0.47 & 0.82 \\
        MEE + QTN & 0.74 & 0.48 & 0.83 \\
        PNC + QTN & 0.59 & 0.41 & 0.72 \\

        MEE + PNC + QTN & 0.75 & 0.49 & 0.84 \\
        QTN$_p$ & 0.60 & 0.41 & 0.73 \\
        MEE + MEE$^2$ & 0.63 & 0.42 & 0.75 \\

        \hline 
    \end{tabular}
\end{table*}

%% file: figures/cmat.tex
\begin{figure*}[ht]
    \centering
    \includegraphics[width=0.48\textwidth]{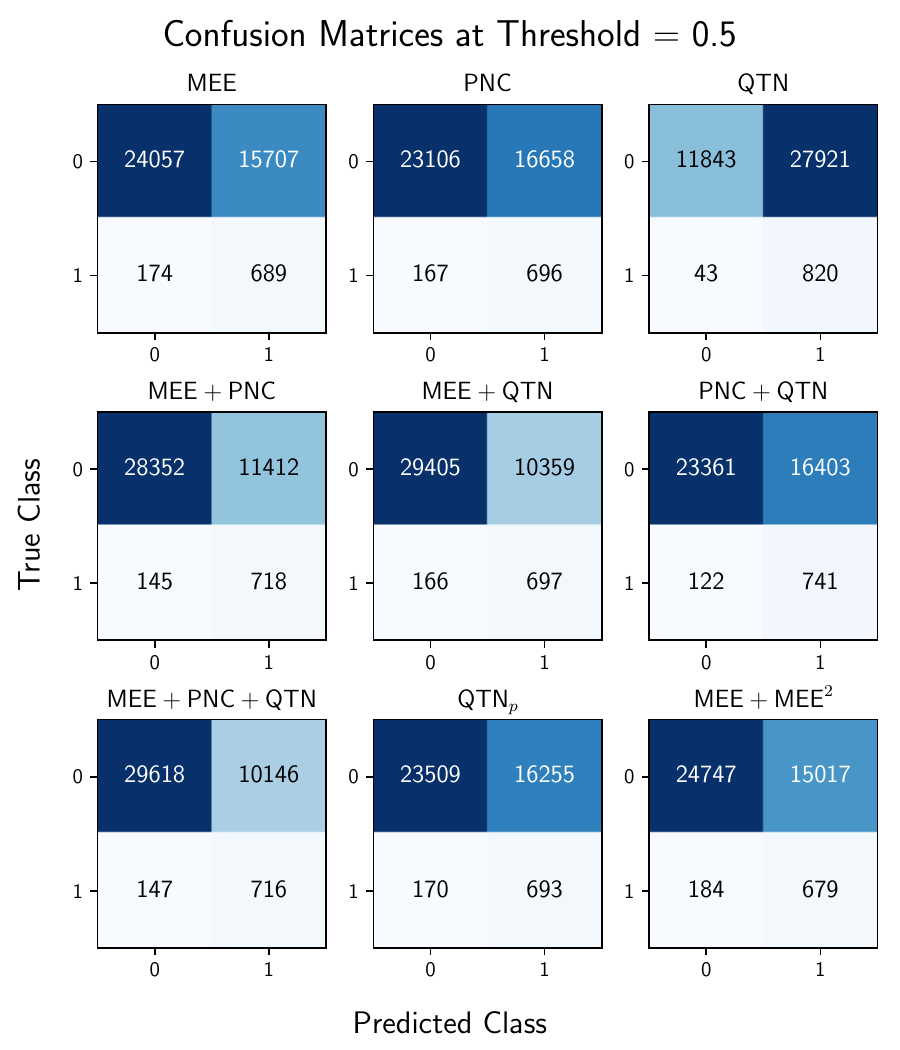}
    \includegraphics[width=0.48\textwidth]{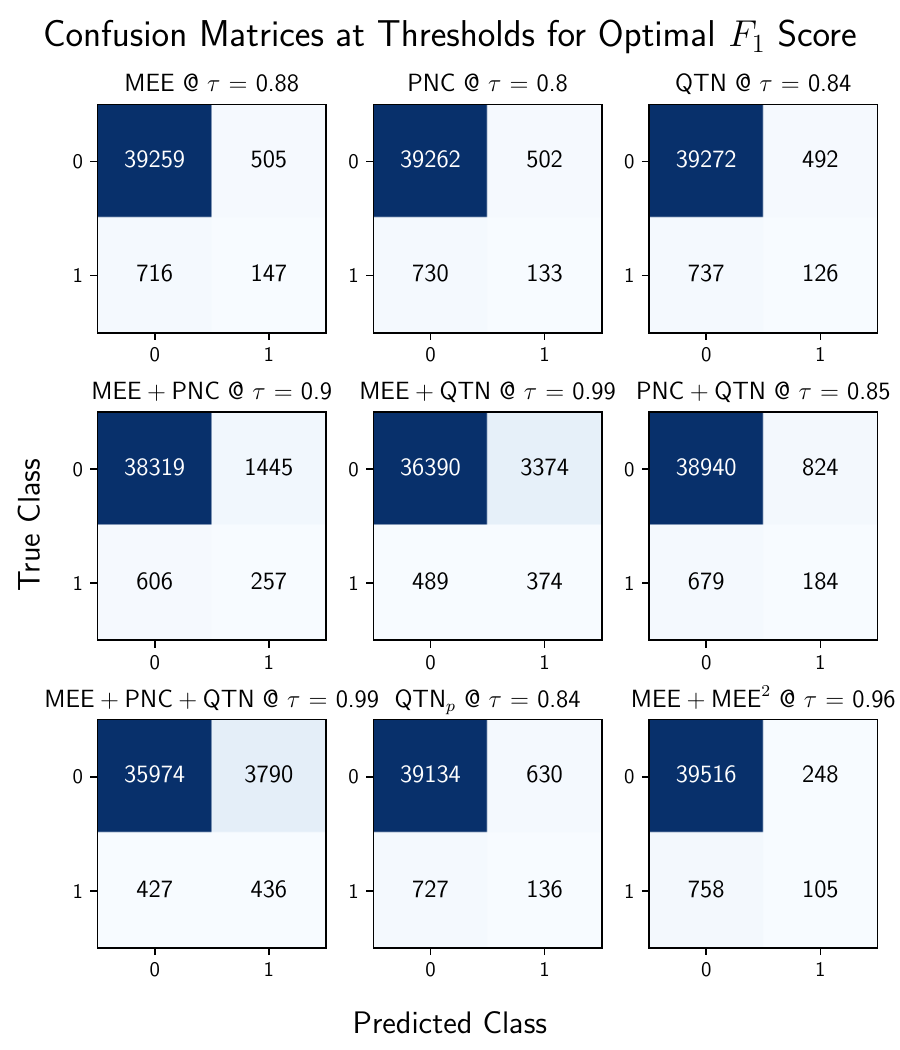}
    \caption{Confusion matrices for each model at default 0.5 threshold, and 
    thresholds selected for optimal macro $F_1$ score. Classes denote whether a
    fragment pair originated from the same parent object; 0 is the negative 
    class, and 1 is the positive class.
    }
    \label{fig:cmat}
\end{figure*}

%% file: figures/roc.tex

\begin{figure}[ht]
    \centering
    \includegraphics[width=0.47\textwidth]{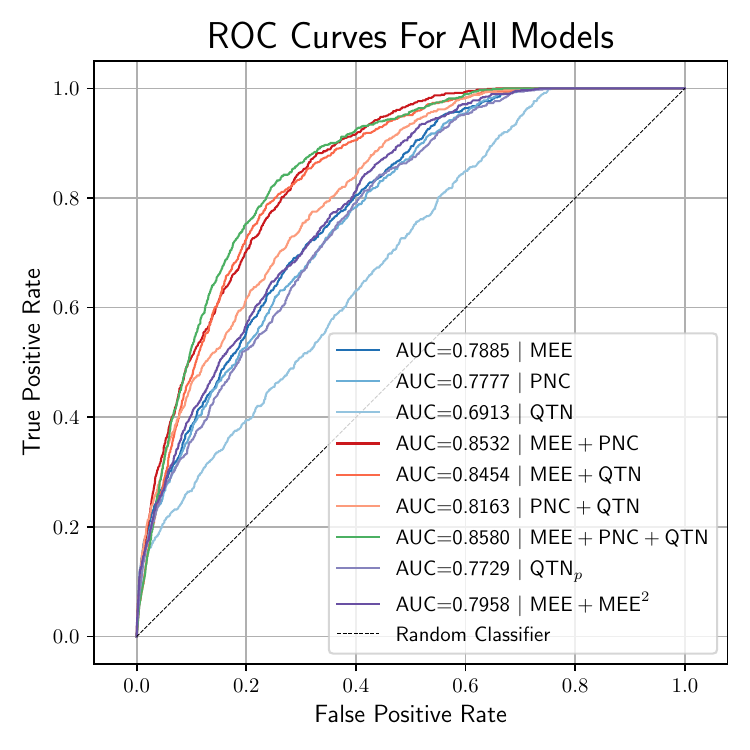}
    \caption{Overview of ROC for all models.
    }
    \label{fig:roc}
\end{figure}

%% file: sections/conclusion.tex
\section{Conclusion}
\label{sec:conclusion}
This paper has presented a modular computational pipeline that synthesizes breakup events of selected parent objects, propagates their fragments under a high-fidelity dynamical model, extracts proper elements from multiple element sets, and applies supervised learning to reconstruct debris families in low Earth orbit. In doing so, it (i) introduces a modular, end-to-end simulation and classification framework for RSOs with unknown origin, (ii) systematically evaluates how proper-element representations and input design affect supervised performance, and (iii) clarifies design choices that improve robustness in crowded, LEO-like environments.

Within this framework, we adapted NASA's EVOLVE 4.0 Standard Breakup Model to generate a large synthetic catalog of fragments from Starlink-like orbits and propagated them for 60 days under major perturbations. Proper elements were then extracted using circle-fitting techniques on modified equinoctial, Poincar\`e, and quaternion element sets, with an interval-sampling inner loop improving robustness over contiguous windows. This produced a clean, labelled dataset of fragment families suitable for controlled evaluation of classification models.

Our experiments show that input representation and element-set design have a substantial impact on performance. A multilayer perceptron trained on absolute proper elements for both fragments outperforms ``lossy" inputs based solely on first- and second-order differences. Adding explicit quadratic terms (MEE$^2$) yields only marginal benefit, indicating that the network can internalize such nonlinearities. In contrast, restoring state completeness in the quaternion set via QTN$_p$ proved critical, significantly enhancing the discriminative power of quaternion elements. This underscores that while multi-set representations offer complementary information, preserving the physical completeness of the underlying state is paramount. The combined element set (MEE + PNC + QTN) yielded the best overall results, achieving an accuracy of 0.75 and a weighted $F_1$ of 0.84 at a fixed decision threshold, along with the highest area under the ROC curve. In this controlled Starlink-like LEO experiment, this represents a clear improvement over our MEE-only baseline and suggests that extending proper-element sets beyond MEE can provide a useful additional degree of freedom when designing RUO classifiers in similar environments. 


Future work must bridge the architectural gap between the supervised classification presented here and the unsupervised clustering required for discovering uncataloged debris families. While this work confirms the discriminative power of multi-set proper elements for known parent associations, operational systems face an "open-set" problem where new breakups lack a priori labels. A robust solution lies in hybrid semi-supervised frameworks that use density-based clustering to propose candidate families and supervised models, such as Graph Neural Networks \cite{Joshi2025}, to validate physical links. Additionally, extending this pipeline to include probabilistic breakup models \cite{Frey2021} and real tracking data will be essential to address the domain shift between synthetic training environments and the chaotic reality of the orbital regime.


%% file: sections/acknowledgements.tex
\section*{Acknowledgement}
We acknowledge that artificial intelligence tools, including OpenAI's ChatGPT, were used in proofreading and polishing this paper.